\documentclass[preprint2]{emulateapj}
\usepackage{multirow}

\newcommand{\reduceme}{\mbox{R\raisebox{-0.35ex}{E}D%
\hspace{-0.05em}\raisebox{0.85ex}{uc}\hspace{-0.90em}%
\raisebox{-.35ex}{{m}}\hspace{0.05em}E}}

\shorttitle{KDC in dEs in the Virgo cluster}
\shortauthors{Toloba et al.}

\begin{document}

\title{Stellar Kinematics and Structural Properties of Virgo Cluster Dwarf Early-Type Galaxies from the SMAKCED Project. I. Kinematically Decoupled Cores and Implications for Infallen Groups in Clusters}

\author{E.~Toloba\altaffilmark{1,2}\footnote{Fulbright Postdoctoral Fellow}}
\email{toloba@ucolick.org}

\author{P.~Guhathakurta\altaffilmark{1}}

\author{G.~van de Ven\altaffilmark{3}}

\author{S.~Boissier\altaffilmark{4}}

\author{A.~Boselli\altaffilmark{4}}

\author{M.~den Brok\altaffilmark{5}}

\author{J.~Falc\'on-Barroso\altaffilmark{6,7}}

\author{G.~Hensler\altaffilmark{8}}

\author{J.~Janz\altaffilmark{9,10}}

\author{E.~Laurikainen\altaffilmark{10,11}}

\author{T.~Lisker\altaffilmark{9}}

\author{S.~Paudel\altaffilmark{12}}

\author{R.~F.~Peletier\altaffilmark{13}}

\author{A.~Ry\'s\altaffilmark{6,7}}

\author{H.~Salo\altaffilmark{10}}

\affil{$^1$UCO/Lick Observatory, University of California, Santa Cruz, 1156 High Street, Santa Cruz, CA 95064, USA}
\affil{$^2$Observatories of the Carnegie Institution for Science, 813 Santa Barbara Street, Pasadena, CA 91101, USA}
\affil{$^3$ Max Planck Institute for Astronomy, K$\ddot{\rm o}$nigstuhl 17, 69117 Heidelberg, Germany}
\affil{$^{4}$ Laboratoire d'Astrophysique de Marseille-LAM, Universit\'e d'Aix-Marseille \& CNRS, UMR 7326, 38 rue F. Joliot-Curie, 13388 Marseille Cedex 13, France}
\affil{$^{5}$ Department of Physics and Astronomy, University of Utah, Salt Lake City, UT 84112, USA}
\affil{$^6$ Instituto de Astrof\'{i}sica de Canarias, V\'{i}a L\'{a}ctea s$/$n, La Laguna, Tenerife, Spain}
\affil{$^7$ Departamento de Astrof\'{i}sica, Universidad de La Laguna, E-38205, La Laguna, Tenerife, Spain}
\affil{$^{8}$ University of Vienna, Department of Astrophysics, T${\rm \ddot{u}}$rkenschanzstra${\rm \ss}$e 17, 1180 Vienna, Austria}
\affil{$^9$ Astronomisches Rechen-Institut, Zentrum f${\rm \ddot{u}}$r Astronomie der Universit${\rm \ddot{a}}$t Heidelberg, M${\rm \ddot{o}}$nchhofstra${\rm \ss}$e 12-14, D-69120 Heidelberg, Germany}
\affil{$^{10}$ Division of Astronomy, Department of Physics, P.O. Box 3000, FI-90014 University of Oulu, Finland}
\affil{$^{11}$ Finnish Center for Astronomy with ESO (FINCA), University of Turku, Finland}
\affil{$^{12}$ Laboratoire AIM Paris-Saclay, CNRS/INSU, Universit\'e Paris Diderot, CEA/IRFU/SAp, 91191 Gif-sur-Yvette Cedex, France}
\affil{$^{13}$ Kapteyn Astronomical Institute, Postbus 800, 9700 AV Groningen, The Netherlands}

\begin{abstract}

We present evidence for kinematically decoupled cores (KDCs) in two dwarf early-type (dE) galaxies in the Virgo cluster, VCC~1183 and VCC~1453, studied as part of the SMAKCED stellar absorption-line spectroscopy and imaging survey. These KDCs have radii of 1.8$''$ (0.14~kpc) and 4.2$''$ (0.33~kpc), respectively. Each of these KDCs is distinct from the main body of its host galaxy in two ways: (1) inverted sense of rotation; and (2) younger (and possibly more metal-rich) stellar population. The observed stellar population differences are probably associated with the KDC, although we cannot rule out the possibility of intrinsic radial gradients in the host galaxy. We describe a statistical analysis method to detect, quantify the significance of, and characterize KDCs in long-slit rotation curve data. We apply this method to the two dE galaxies presented in this paper and to five other dEs for which
KDCs have been reported in the literature. Among these seven dEs, there are four significant KDC detections, two marginal KDC detections, and one dE with an unusual central kinematic anomaly that may be an asymmetric KDC.The frequency of occurence of KDCs and their properties provide important constraints on the formation history of their host galaxies. We discuss different formation scenarios for these KDCs in cluster environments and find that dwarf-dwarf wet mergers or gas accretion can explain the properties of these KDCs. Both of these mechanisms require that the progenitor had a close companion with a low relative velocity. This suggests that KDCs were formed in galaxy pairs residing in a poor group environment or in isolation whose subsequent infall into the cluster quenched star formation.

\end{abstract}

\keywords{galaxies: dwarf -- galaxies: elliptical -- galaxies: clusters: individual (Virgo) -- galaxies: kinematics and dynamics -- galaxies: stellar content -- galaxies: photometry}

\section{Introduction} \label{intro}

Early-type galaxies (ETGs) are characterized by a smooth surface brightness distribution, minimal interstellar gas and dust, and a red color indicating an old stellar population. These galaxies are sometimes referred to as quiescent or quenched, in contrast to star-forming galaxies. The ETG class is composed of two morphological classes, ellipticals (Es) and lenticulars (S0s), distinguished by the absence/presence of a stellar disk.  Dwarf early-type galaxies (dEs for simplicity)\footnote{The term dE has traditionally been used to refer to dwarf elliptical galaxies, whereas we loosely use the term here to include dwarf elliptical, dwarf lenticular (dS0), dwarf spheroidal (dSph), and even the occasional low-luminosity elliptical (E) galaxies.} occupy the low luminosity ($M_B \gtrsim -18$) and low surface brightness ($\mu_B \gtrsim 22$~mag~arcsec$^{-2}$) section of the ETG class.

Detailed photometric and spectroscopic studies of dEs suggest that they are a heterogeneous population. Beneath their mostly smooth light distribution and apparent morphological simplicity, some dEs display subtle signs of structural complexity in the form of disks, spiral arms, and/or irregular features \citep[e.g.,][]{Jerjen00,Barazza02,Geha03,Graham03,DR03,Lisk06a,Ferrarese06,Janz12}. Kinematically too, dEs display a range of diversity in that galaxies with similar photometric properties can have very different rotation speeds \citep[][]{Ped02,SimPrugVI,Geha02,Geha03,VZ04,Chil09,etj09,etj11,Rys13}. 

The dE galaxy class is dominant by number in high density environments \citep{Sand85,Bing88} and these galaxies are rarely seen in isolation \citep{Gavazzi10}. In a recent study, \citet{Geha12} found that quenched galaxies with stellar masses of $\sim 10^9 M_{\odot}$ and magnitudes of $M_r > -18$ (properties that include those of dEs) are invariably located within 1.5~Mpc ($\sim 4$ virial radii) of a relatively massive host, while star-forming galaxies of similar masses are found beyond that distance. It should be noted though that an exception to the Geha et~al.\ finding is the discovery by \citet{Pasquali05} of a quiescent dwarf at a distance of 1.9~Mpc from its nearest neighbor.

This strong morphological segregation and their structural complexity suggest that dEs are late-type galaxies transformed by the environment. A few different transformation mechanisms have been proposed over the years, but none of these mechanisms appear to explain all of the observed properties of dEs. The most promising of these mechanisms involve gravitational tidal heating \citep[i.e., harassment, as described in][]{Moore98,Mast05} or hydrodynamical interactions with the intracluster medium \citep[i.e., ram-pressure stripping, as discussed in][]{LinFab83,Boselli08a,Boselli08b}.

With the goal of understanding the physical processes that form dEs, we have begun the SMAKCED\footnote{http://smakced.net/index.html} (Stellar content, MAss and Kinematics of Cluster Early-type Dwarf galaxies) project, a new spectroscopic and photometric survey of dEs in the Virgo cluster, the nearest dense galaxy cluster. This paper is part of a series in which we present analysis of the structural and kinematical properties of Virgo cluster dEs. Details of the photometric and spectroscopic data sets are described in Toloba et~al.\ (2014, in preparation --- hereafter referred to as Paper~II). This paper, the first in the series, is focused on the analysis of newly discovered kinematically decoupled cores (KDCs) in two of the dEs in our sample, VCC~1183 and VCC~1453.

A KDC refers to a region in the center of a galaxy whose kinematical properties are distinct from those of the main body of the galaxy. While KDCs have long been known in luminous ETGs \citep[e.g.,][]{Rix92,SurmaBender95,Mehlert98,Wernli02,Em04,Geha05,Em11,Kraj11}, they are harder to detect in dEs due to their lower surface brightness. Nevertheless, it is important to study and characterize KDCs in dEs in order to understand the physical processes that produce them, including how these processes may be distinct from those that produce KDCs in luminous ETGs. 

The formation of KDCs in luminous ETGs is thought to be associated with gas accretion, merging scenarios, or flyby encounters \citep[e.g.,][]{Balcells90,Hernquist91,Weil93,GonzalezGarcia05}. Whether these scenarios apply to dEs is far from obvious given that dEs are mainly found in galaxy clusters where there is a dearth of cold gas to accrete, the statistical probability of dwarf-dwarf mergers is low (e.g., as indicated by the simulations and observations of \citet{deLucia06} and \citet{MartinezDelgado12}, respectively), and flyby encounters between cluster members are so rapid that the energy available is too small to produce such perturbations in the core of a dwarf galaxy \citep{BinnTrem87,BG06}.

This paper is organized as follows. In Section~\ref{obs}, we describe the spectroscopic and photometric data used to find the KDCs in VCC~1183 and VCC~1453. In Section~\ref{tech}, we present the measurement of stellar kinematics and stellar population parameters. In Section~\ref{kin}, we use these measurements to infer the properties of the KDCs and their host galaxies. In Section~\ref{quantKDC}, we introduce a statistical metric to quantify the significance of the detection of the two new KDCs as well as other candidate KDCs from the literature. In Section~\ref{disc}, we discuss the most likely formation scenarios for KDCs. Finally, in Section~\ref{concl}, we  summarize our findings and conclusions.

\section{The Data} \label{obs}

This paper uses $H$-band photometry and optical spectroscopy collected as part of the SMAKCED project. We present a brief summary here of the sample selection, observations, and data reduction. The reader is referred to Paper~II for more details.

\subsection{The Sample}

The photometric SMAKCED survey, conducted in the $H$-band, consists of a nearly complete sample of 121 early-type galaxies in the Virgo cluster in the absolute $r$-band magnitude range $-19.0 < M_r < -16.0$, including a complete magnitude-limited subsample with $M_r < -16.8$. We assume a distance to the Virgo cluster of 16.5 Mpc \citep{Mei07}. The galaxies are selected from the Virgo Cluster Catalog \citep[VCC,][]{Bing85} based on their early-type morphology and low luminosity, see details in \citet{Janz13}.
While this sample consists mostly of galaxies with dwarf elliptical or dwarf lenticular (dS0) morphological classifications in the literature, it also contains a small fraction (9.9\%, 12/121) of galaxies with elliptical (E) morphological classification. For the purpose of this series of papers, we loosely refer to all of these ETGs in this magnitude range as dEs.

The spectroscopic survey consists of a subsample of 39 dwarf early-type galaxies in the absolute $r$-band magnitude range $-18.6<M_r<-16.4$ classified as dwarf elliptical or dwarf lenticular (dS0) by \citet{Bing85}\footnote{Low-luminosity ETGs with high surface brightness are not included in the spectroscopic subsample.}. These 39 dEs are representative of the Virgo cluster ETGs in the magnitude range $-19.0 < M_r < -16.0$. More details on the spectroscopic sample can be found in \citet[][]{etj11} and Paper~II.

\begin{table}
\begin{center}
\caption{Galaxy Properties \label{props}}
\resizebox{9cm}{!} {
\begin{tabular}{|c|c|c|c|c|c|c|}
\hline
Galaxy  &  RA(J2000) & Dec(J2000) & $m_r$  &  $m_H$  &  $R_e$ ($''$)  &  $n$ \\
(1)        &      (2)         &   (3)            &  (4)      &     (5)      &    (6)              & (7)    \\
\hline \hline

VCC~1183 & 12:29:22.5   & $+$11:26:02  &  13.27 & 11.12   &  19.1             &  1.72  \\
VCC~1453 & 12:32:44.2   & $+$14:11:44 & 13.25 & 11.00   &  16.9             &  2.22  \\

\hline
\end{tabular}
}
\end{center}
NOTES: (1) galaxy name; (2) right ascension in hh:mm:ss; (3) declination in dd:mm:ss; (4) magnitude in the $r$-band by \citet{Janz08}; (5) magnitude in the $H$-band; (6) half-light radius in the $H$-band; (7) S\'ersic index of a single model fit to the surface brightness profile in the $H$-band. Columns (5), (6), and (7) are measurements by \citet{Janz13}.
\end{table}

This paper focuses on the two dEs, out of the 39 analyzed in Paper~II, that turned out to host a KDC: VCC~1183 and VCC~1453. Some of their basic properties can be found in Table~\ref{props}. 

\subsection{Observations and Data Reduction}

The observations and data reduction for the $H$-band photometry are detailed in \citet{Janz13}, the optical spectroscopy  is detailed in \citet{etj11} and Paper~II. Here we summarize the instrumental setups and the main steps of the reduction procedure.

VCC~1183 was observed for 720\,s using the SOFI infrared camera at the 3.6-m New Technology Telescope (NTT) at the ESO Observatory (Chile). VCC~1453 was observed for 2820\,s using the NOTCam wide-field camera at the 2.5-m Nordic Optical Telescope (NOT) at El Roque de los Muchachos Observatory (Spain).
The observations were done using standard dither patterns that allowed us to get the sky level simultaneously with the target galaxy. 

The data reduction was performed with IRAF\footnote{IRAF is distributed by the National Optical Astronomy Observatory, which is operated by the Association of Universities for Research in Astronomy, Inc., under cooperative agreement with the National Science Foundation.} following the standard steps for infrared photometry. These included illumination correction, flat fielding, sky subtraction, and correction for field distortions.

The spectroscopic observations were performed at El Roque de los Muchachos Observatory (Spain).

VCC~1183 was observed as part of the MAGPOP-ITP collaboration (Multiwavelength Analysis of Galaxy POPulations-International Time Program, PI: R.~Peletier) and already presented in \citet{etj11}. In this paper, we reanalyze the kinematics changing the radial binning scheme from a minimum of 1 pixel to 3 pixels (see Section \ref{tech}). This galaxy was observed for 3600\,s with the 2.5-m Isaac Newton Telescope (INT) using the IDS spectrograph with the 1200\,l/mm grating covering the wavelength range 4600--5690~\AA. 

VCC~1453, shown for the first time here, was observed for 7000\,s with the 4.2-m William Herschel Telescope (WHT) and the double-arm spectrograph ISIS in two independent instrumental setups. The blue setup used the 1200\,l/mm grating and covered 4200--5000~\AA. The red setup used the 600\,l/mm grating and covered 5500--6700~\AA. 

Both galaxies were observed using a slit width of 2$''$ which leads to a spectral resolution of 1.6~\AA~ (full width at half the maximum, FWHM) for VCC~1183, and 1.4 and 3.2~\AA~ (FWHM) for VCC~1453 in the blue and red setup, respectively. 

The data was reduced following the standard procedure for long-slit spectroscopy using the package \reduceme~ \citep{Car99}. The main steps consisted of bias and dark current subtraction, flat fielding, and cosmic ray cleaning. The spectra were spatially aligned and wavelength calibrated leading to typical wavelength residuals of 0.2~\AA. The spectra were then sky subtracted and flux calibrated using the response function derived from our observed flux standards.

\section{Measurement of the Kinematic and Stellar Population Profiles}\label{tech}

From the reduced two dimensional spectra, we measure the line-of-sight radial velocity ($V$) and velocity dispersion ($\sigma$) along the galaxy's major axis using the penalized pixel-fitting software (pPXF) of \citet{PPXF}. 
The parameters $V$ and $\sigma$ are obtained by cross-correlating each science spectrum with a model created as a linear combination of stellar templates that best reproduces the science spectrum, allowing different weights for each template.
The stellar templates used for this analysis are a compilation of stars selected from the MILES stellar library \citep{SB06lib,MILEScen,FB11MILES} and cover a variety of spectral types and luminosity classes (B9, A0, A5V, G2III, G2V, G8III, G9III, K0I, K1V, K2III, K3III, K4III, M2III). These templates were observed with the same instrumental setup as the galaxies. We defocused the telescope to make the stars homogeneously fill the slit as galaxies do. 

Each one dimension science spectrum is obtained by spatially co-adding individual spectra following the size and S/N thresholds described in \citet{etj11} and in Paper~II. The minimum binning size is 3 pixels (comparable to the average seeing affecting the observations), and the minimum S/N is 10~\AA$^{-1}$ for $V$ and 15~\AA$^{-1}$ for $\sigma$. The minimum S/N threshold for the red configuration of VCC~1453 is ${\rm S/N} = 15$ and ${\rm S/N} = 25$~\AA$^{-1}$ for $V$ and $\sigma$, respectively, to compensate for the several masked regions due to sky line residuals.

\begin{figure*}
\centering
\includegraphics[angle=-90,width=12cm]{VCC1183_KDC.ps}\\
\includegraphics[angle=0,width=4cm,bb= 167 750 453 100]{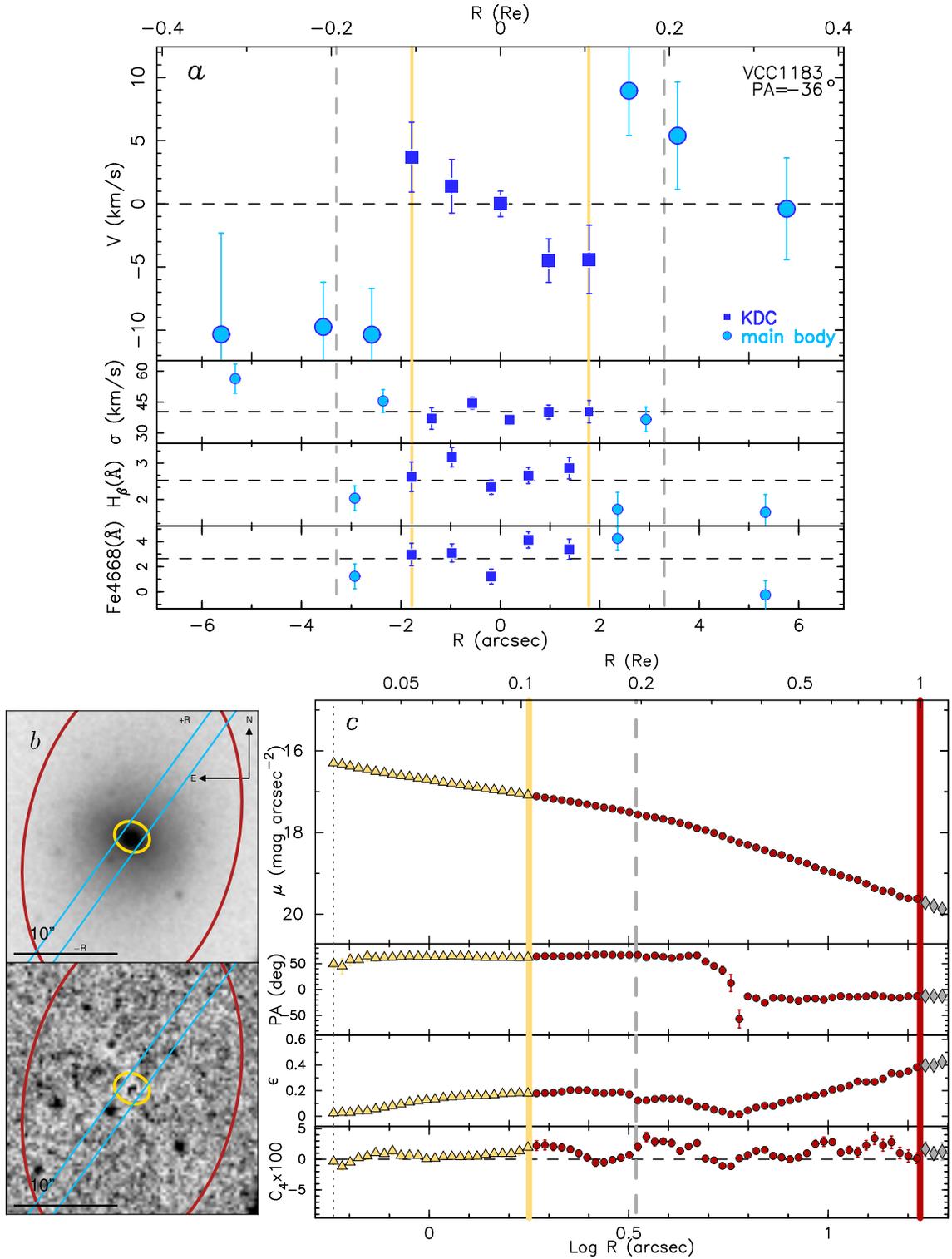}
\includegraphics[angle=-90,width=11cm]{VCC1183_profiles_log.ps}
\caption{Spectroscopic and photometric measurements of VCC~1183. \textbf{Panel \textit{a}:} Stellar kinematics and population indices as a function of distance from the galaxy's center along the slit (in units of $R_e$ and arcseconds on the upper and lower horizontal axes, respectively). The slit position angle is indicated in the upper right and corresponds to the part of the slit with positive radial distances. The rows, from top to bottom, show the stellar rotation curve (with respect to the systemic velocity), the stellar velocity dispersion profile, the H$_{\beta}$ index as a tracer of age, and the Fe4668 index as a tracer of metallicity. The filled dark blue squares indicate the kinematically decoupled core based on the rotation curve and the light blue circles show the main body of the galaxy. The yellow vertical lines indicate the radii for the maximum rotation of the KDC ($R_{\rm vmax}^{\rm KDC}$). \textbf{Panel \textit{b}:} Zoom on the central regions of the $H$-band image (upper panel) and residuals (lower panel) after subtracting a smooth model based on ellipse-fitting. The grey scales show bright regions in black. The blue lines are the footprint of the long-slit used in the spectroscopic observations. The $+R$ and $-R$ correspond to the positive and negative radial distances measured in the spectroscopy with respect to the center of the galaxy. The yellow and red ellipses are the corresponding elliptical isophotes with semi-major axes of $R_{\rm vmax}^{\rm KDC}$ and $R_e$, respectively. \textbf{Panel \textit{c}:} Profiles of the ellipse-fitting. From top to bottom: surface brightness, $PA$, ellipticity, and $C_4$ profiles. The grey dotted line indicates a radius of 2 pixels (${\rm 1~pixel}=0.288''$). The ellipse-fitting results are unreliable interior to this radius and are therefore excluded from the plots. The yellow vertical lines, enclosing the triangles, are the same as in panel \textit{a}. The red vertical lines, enclosing circles, indicate the $R_e$. The dashed grey line indicates the radius at which the inner S\'ersic and outer exponential profiles fitted to the $H$-band images reach the same surface brightness \citep{Janz13}. 
\label{gradients_VCC1183}}
\end{figure*}

\begin{figure*}
\centering
\includegraphics[angle=-90,width=12cm]{VCC1453_KDC.ps}\\
\includegraphics[angle=0,width=4cm,bb= 167 750 453 100]{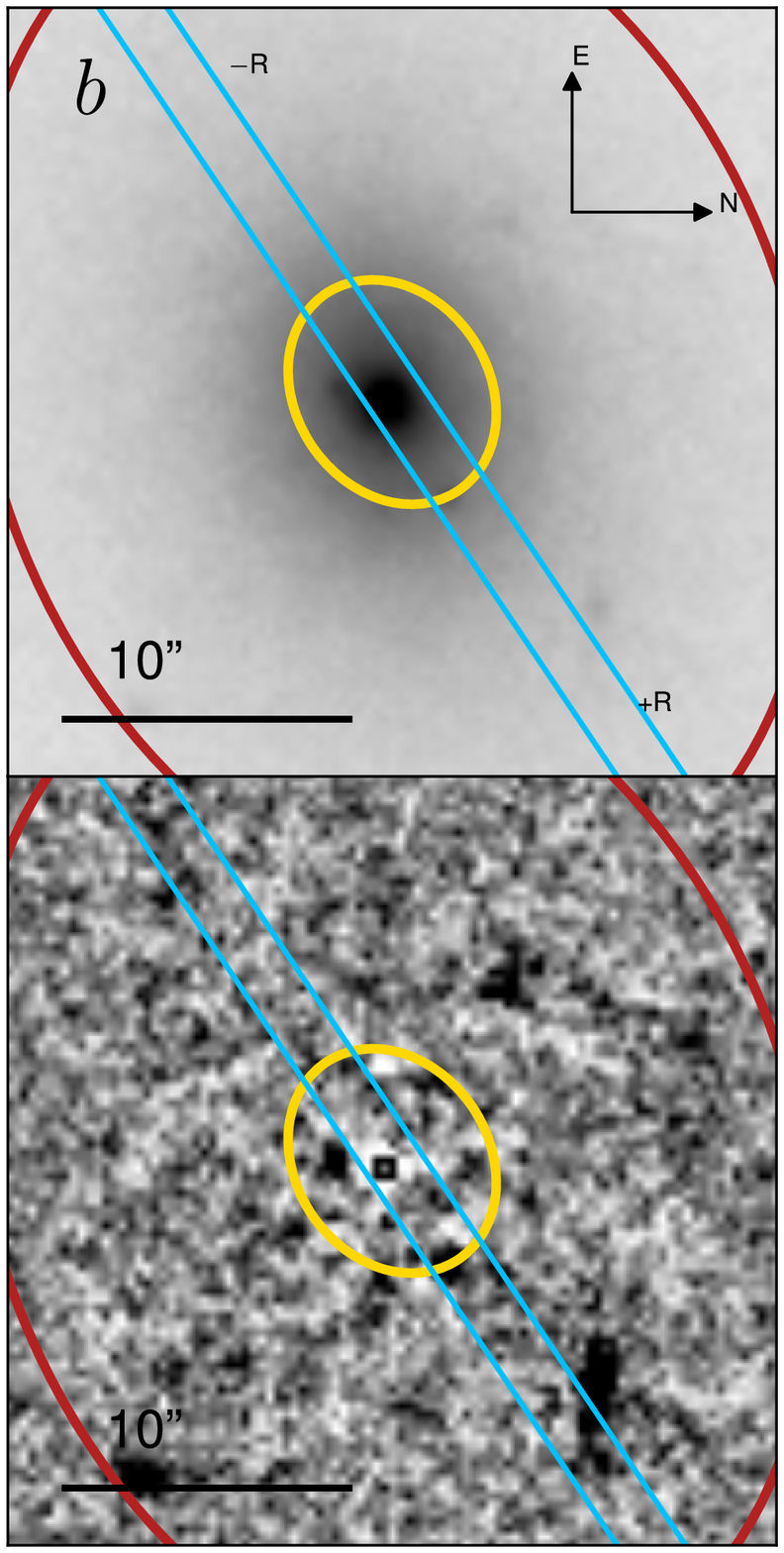}
\includegraphics[angle=-90,width=11cm]{VCC1453_profiles_log.ps}
\caption{Same as Figure \ref{gradients_VCC1183} for VCC~1453, except the metallicity tracer used for this galaxy is Fe5709 and the pixel scale is $0.234''$.
\label{gradients_VCC1453}}
\end{figure*}

The resulting kinematical profiles are shown in the first two rows of panel {\it a} in Figures \ref{gradients_VCC1183} and \ref{gradients_VCC1453}. The rotation curve shown in the first row is obtained subtracting to each radial velocity the systemic velocity which is the weighted average of all the velocities along the slit. This technique finds the center of the amplitude of the rotation curve. The second row shows the velocity dispersion profile.

As VCC~1453 was observed with two independent instrumental setups, the final rotation curve is a weighted average of the velocity measurements obtained from the blue and red setups in those regions where they overlap. At larger radii, only the red setup is used. Regarding the $\sigma$ profile, we take advantage of the higher resolution of the blue setup and use only those values.

The stellar populations along the major axis of the galaxies are estimated using hydrogen (H$_{\beta}$) and iron (Fe4668 and Fe5709) Lick spectral indices \citep{Wor94}. H$_{\beta}$ is used  as an indicator of age, and Fe4668 and Fe5709 are used as indicators of metallicity. We use Fe5709 for VCC~1453 to have the largest radial coverage possible. 

These Lick indices are measured in the Line-Index-System LIS-5\AA, i.e., broadened to a constant FWHM of 5\AA~\citep{Vazdekis10}. The spatial co-addition of the spectra to measure the spectral indices follows the same strategy as that for $V$ and $\sigma$ with the exception that the minimum binning size required is 4 pixels to have larger S/N per spatial bin.

The last two rows of panel {\it a} in Figures \ref{gradients_VCC1183} and \ref{gradients_VCC1453} show the resulting H$_{\beta}$ and Fe4668, for VCC~1183, and Fe5709, for VCC~1453, profiles.

\section{Analysis of the Properties of the KDCs} \label{kin}

In this Section we explore whether the KDCs, identified in velocity space, have properties that are different from those of the main body of the host galaxy. In addition to the rotation velocity, we investigate the velocity dispersion profile, the stellar populations, and the structure of these galaxies.

\subsection{Kinematics}

The rotation curves of VCC~1183 and VCC~1453 show cores with inverted sense in velocity with respect to the main body of the galaxy. These KDCs have radial sizes of 1.8$''$ (0.14~kpc) and 4.2$''$ (0.33~kpc), respectively. 

The velocity dispersion profile of VCC~1183 is rather flat, thus the KDC does not show a different velocity dispersion from that of the main body of the galaxy. The velocity dispersion profile of VCC~1453  has a dip in the center of the galaxy with respect to the values at larger radii. We do not know how the velocity dispersion of the main body of the galaxy compares with that of the KDC because the data do not go beyond the radial extent of the KDC.

The first and second rows of panel {\it a} in Figures \ref{gradients_VCC1183} and \ref{gradients_VCC1453} show the rotation curve and velocity dispersion profile for VCC~1183 and VCC~1453, respectively. The dark blue squares highlight the KDCs. The yellow vertical lines indicate the radius at which the velocity of the KDC reaches its maximum ($R_{\rm vmax}^{\rm KDC}$). That radius is calculated averaging the position of the peak velocity for the approaching and receding radii. \citet{Janz13} make a two dimensional fit  to the $H$-band surface brightness distribution of these dEs. The dashed grey lines indicate the radius at which the inner S\'ersic and outer exponential profiles have the same surface brightnesses. 

\subsection{Stellar Populations}

Assuming the single stellar population models (SSPs) of \citet{Vazdekis10} based on the MILES stellar library \citep{SB06lib,MILEScen,FB11MILES} with a Kroupa initial mass function \citep{KroupaIMF}, we estimate the ages and metallicities ([M/H]) of the KDC and the main body of the galaxy using the software {\sc rmodel\footnote{\url{http://www.ucm.es/info/Astrof/software/rmodel/rmodel.html}}} \citep[]{Cardiel03}. This software interpolates the age and metallicity inside an index-index grid. The errors in the age and [M/H] are calculated running 1000 Monte Carlo simulations varying the values of the spectral indices within a Gaussian distribution with width equal to their uncertainties. 

The KDC of VCC~1183 is $2.0\pm0.5$~Gyr old and has a metallicity of ${\rm [M/H]} = -0.3\pm0.3$, while the main body is $\gtrsim 10$~Gyr old (this value is outside the grid so we assume the age in the models that is consistent with the $2\sigma_G$\footnote{Gaussian width, $\sigma_G$} uncertainty of the measurement) with ${\rm [M/H]} = -1.1\pm0.3$. 

The KDC of VCC~1453 is $4.5\pm1.1$~Gyr old, and has a metallicity of ${\rm [M/H]} = -0.2\pm0.1$. The main body of the galaxy has an age of $10.3\pm5.7$~Gyr and a metallicity of ${\rm [M/H]} = -0.6\pm0.3$. 

The stellar populations for VCC~1183 and VCC~1453 are younger and possibly more metal-rich in the KDC than in the main body of the galaxy.
In general, having younger and more metal-rich central regions is not unusual for nucleated dEs  \citep{Paudel11}. However, only in extreme cases does this trend go beyond the nuclear region ($R \sim 1''$).

\begin{figure*}
\centering
\includegraphics[angle=-90,width=8cm]{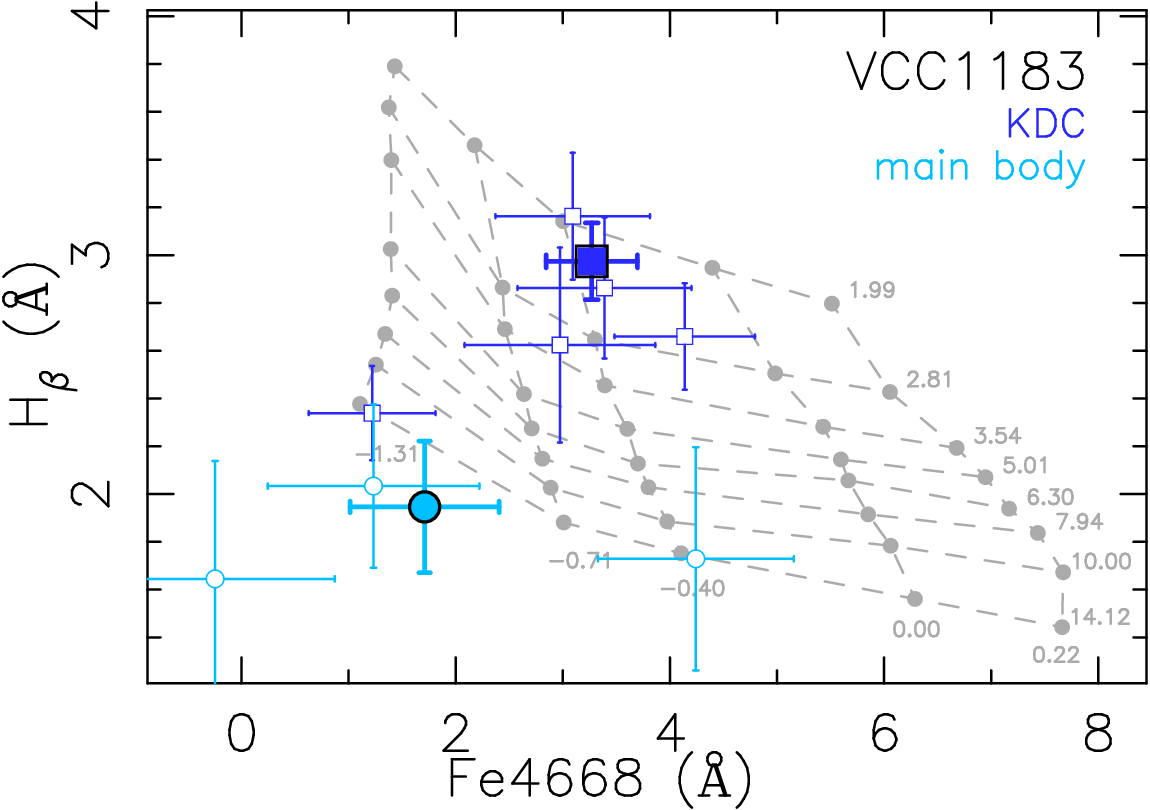}
\includegraphics[angle=-90,width=8cm]{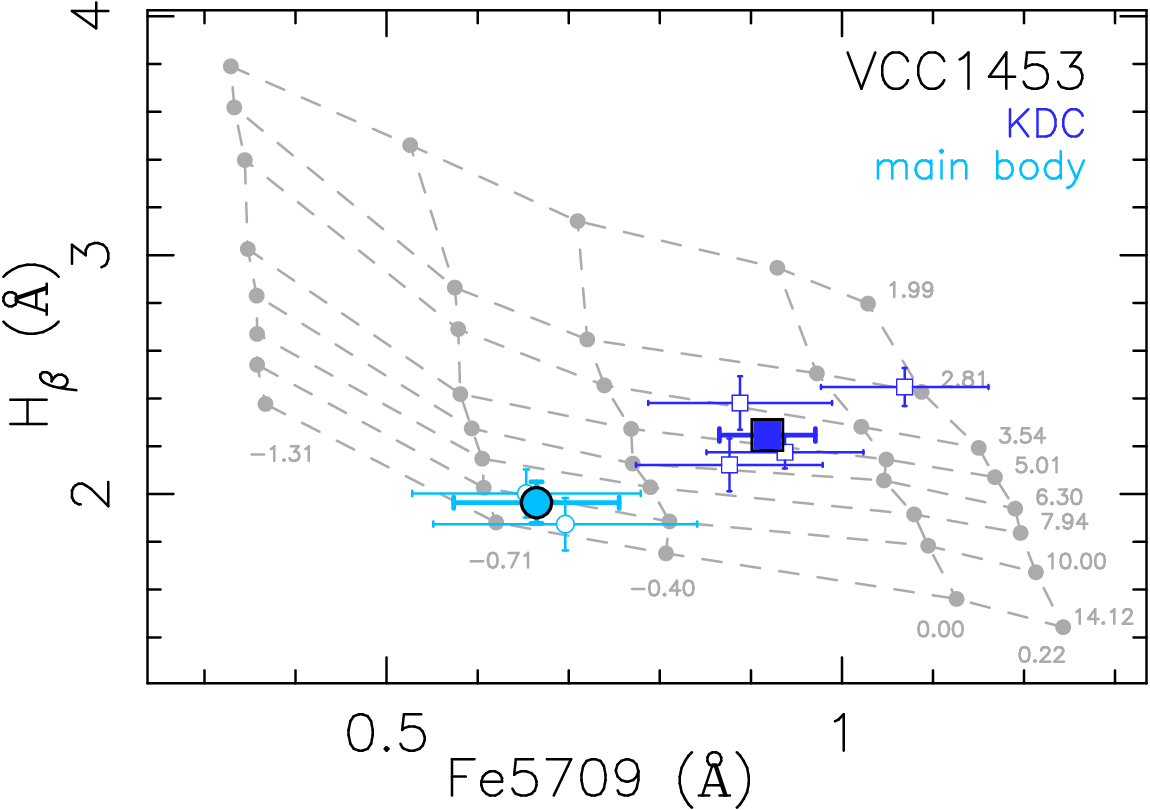}
\caption{Spectral index-index diagrams used to estimate the stellar populations of galaxies. The grey dashed lines represent the grid of models by \citet{Vazdekis10} in the system LIS-5\AA~ based on the MILES stellar library \citep{SB06lib,MILEScen,FB11MILES} with a Kroupa initial mass function \citep{KroupaIMF}. Nearly horizontal lines indicate constant ages, values in Gyr printed at the right end of those lines, and nearly vertical lines show constant metallicity, labeled in the lower part of the grid. Following the color code from Figures \ref{gradients_VCC1183} and \ref{gradients_VCC1453}, the open dark blue squares correspond to the KDC, and the light blue dots represent the main body of the galaxy. The solid dark blue square and light blue circle are the resulting spectral indices after the co-addition of the spectra inside and outside of the KDC region.
\label{indices}}
\end{figure*}

The last two rows of panel {\it a} in Figures \ref{gradients_VCC1183} and \ref{gradients_VCC1453} show the radial profiles of the spectral indices used as age and metallicity indicators for VCC~1183 and VCC~1453.
Figure \ref{indices} shows the spectral index-index diagrams used to determine the stellar populations. The open dark blue squares and light blue circles correspond to the H$_{\beta}$ and iron measurements in Figures \ref{gradients_VCC1183} and \ref{gradients_VCC1453}. For VCC~1453, the radial coverage of  H$_{\beta}$ and Fe5709  is different, thus, for each H$_{\beta}$ value we plot the nearest value of Fe5709 in radius. The solid dark blue square and light blue circle are the $H_{\beta}$ and iron indices of the KDC and the main body of the galaxy, respectively. These indices are measured by co-adding the spectra inside and outside of the KDC region. 

\subsection{Photometry}

VCC~1183 and VCC~1453 are structurally complex.  The $H$-band surface brightness profiles of these two galaxies are fitted by several components for the main body of the galaxies and a central nucleus \citep{Janz12,Janz13}. 

We use the IRAF task {\sc ellipse} to fit isophotes to the $H$-band images. Each isophote is an ellipse whose major axis is $5\%$ larger than the previous one. The center, the position angle ($PA$), and the ellipticity ($\epsilon$) of the isophotes are left as free parameters.

The analysis of the images and the resulting profiles from ellipse-fitting do not show any significant structure that can be associated with the KDC. However, VCC~1183 shows a particularly complex structure that includes the KDC and the remaining region with kinematical data. At $R \sim 6''$ the isophotes twist $\sim 77^{\circ}$  and the ellipticity drops with respect to the outer parts of the galaxy. The surface brightness distribution at $R \lesssim 6''$ is best fitted by a lens \citep{Janz13}. 

Panel {\it b} of Figures \ref{gradients_VCC1183} and \ref{gradients_VCC1453} shows the $H$-band image of the galaxy (upper panel) and the residuals after subtracting a smooth model based on the ellipse-fitting of the isophotes (lower panel). 
Panel {\it c} shows the profiles for surface brightness, position angle, ellipticity, and $C_4$ parameter ($C_4 < 0$ indicates boxy isophotes and $C_4 > 0$ disky isophotes) obtained by the ellipse-fitting procedure described above.

\section{Detection of KDCs and their Rate of Occurrence}\label{quantKDC}

In this section, we develop a method to objectively define a KDC and assess its statistical significance. We apply this technique to the two new KDCs presented in this paper and also to five other dEs for which KDCs have been reported in the literature. Appendix \ref{app} contains relevant plots for each of the seven dE galaxies.

In addition, we use the results from our statistical method to calculate the rate of occurrence of KDCs in dEs and compare it with the rate of occurrence of KDCs in more luminous early-type galaxies.

\subsection{Development and Application of a Statistical Metric to Detect KDCs}\label{quantKDC1}

We propose a method to quantify whether the detection of a KDC in long-slit spectroscopic data is statistically significant. It is based on a statistical parameter $S(R)$ defined as

\begin{equation}
S(R) = \frac{\vert \langle V_{\rm inner}(<R) \rangle - \langle V_{\rm outer}(>R) \rangle \vert}{\sqrt{\delta \langle V_{\rm inner}(<R) \rangle^2+ \delta \langle V_{\rm outer}(<R) \rangle^2}}
\end{equation}

For each radius $R$, $\langle V_{\rm inner}(<R) \rangle$ is the weighted mean velocity interior to that radius and $\langle V_{\rm outer}(>R) \rangle$ is the weighted mean velocity exterior to that radius. The parameters $\delta \langle V_{\rm inner}(<R) \rangle$ and $\delta \langle V_{\rm outer}(<R) \rangle$ are the uncertainties in the weighted mean velocities, respectively. 

We define the significance of the detection of a KDC ($\langle S_{{\rm max}} \rangle$) as the average of the three maximum values of $S(R)$. The average of their radii is the parameter $\langle R_{S{\rm max}} \rangle$. It is important to distinguish between a monotonically increasing rotation curve from a rotation curve with a KDC in the center.  If a KDC is present, the mean velocity of the KDC has the opposite sign with respect to the mean velocity of the main body of the galaxy. Thus, we calculate $\langle S_{{\rm max}} \rangle$ in the region where $\langle V_{\rm inner}(<R) \rangle$ and $\langle V_{\rm outer}(>R) \rangle$ have opposite signs. A KDC is considered significant when $\langle S_{\rm max} \rangle \gg 3$ and marginal when $\langle S_{\rm max} \rangle \sim 3$.

Table \ref{tableKDCs} shows the results of the application of our statistical analysis to the folded rotation curves of the two KDC candidates found in our sample of dEs and the five other candidates reported in the literature. Four out of the seven KDC candidates meet our thresholds for statistically significant detections and two more are marginal detections. The remaining galaxy, VCC~1036, is different from the other galaxies (see Figure \ref{appVCC1036}). The rotation curve of VCC~1036 is asymmetric in two ways: (1) one side is rotating more rapidly than the other; and (2) one side is kinematically regular in the center while the other side is not. The statistical parameter $S(R)$ for this galaxy has only three values where $\langle V_{\rm inner}(<R) \rangle$ and $\langle V_{\rm outer}(>R) \rangle$ have opposite signs. There is not a clear peak among these three values. Thus, while there is a statistically significant anomaly in the center of this galaxy, it does not resemble a symmetric KDC.

The inconvenience of using long-slit spectra is that the KDC can be missed if it is misaligned with respect to the rotation field of the main body of the galaxy. Thus, not meeting the significance threshold described above does not rule out the presence of a KDC in that dE. This caveat is better addressed with integral-field spectroscopy, but large samples of dEs observed with this technique are still lacking.

\begin{table*}
\begin{center}
\caption{Statistical significance of the detection of KDCs \label{tableKDCs}}
\resizebox{18cm}{!} {
\begin{tabular}{|c|c|c|c|c|c|c|}
\hline \hline
Galaxy  & $\langle R_{S{\rm max}} \rangle$   &  $\langle S_{{\rm max}} \rangle$ &  $\langle V^{\rm KDC}_{{\rm inner}} \rangle$ & $\langle V^{\rm dE}_{{\rm outer}} \rangle$ & KDC?  &  Source\\
            &    arcsec &                 &  km~s$^{-1}$  & km~s$^{-1}$ &       &      \\
 (1)       &   (2)   &     (3)              & (4)         &     (5)              &   (6)       &  (7)    \\
\hline
VCC~1183 & 2.2 $\pm$ 0.4 & 4.6 $\pm$ 0.8 &  $-3.5 \pm$ 1.1 &  7.4 $\pm$ 1.7 &  significant &  This work \\
\hline
VCC~1453 & 12.4 $\pm$  1.4 & 3.2 $\pm$ 0.2 & $-5.4 \pm$ 0.6 &  7.4 $\pm$ 3.9 & marginal &  This work \\
\hline \hline
VCC~510 & 3.3 $\pm$ 1.1 & 4.3 $\pm$ 0.9 & $-3.5 \pm$ 1.4 &  6.8 $\pm$ 1.3 &  significant &  Th06 \\
\hline
VCC~917 & 3.0 $\pm$ 1.1 & 3.3 $\pm$ 0.2 &  $-0.6 \pm$ 0.9 &  5.2 $\pm$ 1.7 &  marginal &  Ch09 \\
\hline
VCC~1036$^*$ & 1.0 $\pm$ 0.3 & 12.5 $\pm$ 3.2 & $-0.2 \pm$ 1.0 &  13.7 $\pm$ 0.6 &  ? &  Ch09 \\
\hline
VCC~1261 & 9.4 $\pm$ 0.9 & 5.6 $\pm$ 0.3 & $-2.79 \pm$ 0.6 &  6.4$\pm$ 1.6 &  significant &  Ch09 \\
\hline
VCC~1475 & 1.7 $\pm$ 0.2 & 56.9 $\pm$ 0.2 & $-1.25 \pm$ 0.1 &  6.5 $\pm$ 0.1 & significant & Ko11 \\
\hline
\end{tabular}
}
\end{center}
NOTES: Column 1: Galaxy name; Column 2: average radius of the three maximum values of $S(R)$ where $\langle V_{\rm inner}(<R) \rangle$ and $\langle V_{\rm outer}(>R) \rangle$ have opposite signs. This parameter indicates the radius where the rotation curve goes from negative velocity values to positive velocity values; Column 3: significance of the KDC, i.e., average of the three maximum values of $S(R)$ where $\langle V_{\rm inner}(<R) \rangle$ and $\langle V_{\rm outer}(>R) \rangle$ have opposite signs; Column 4: average velocity in the region interior to $\langle R_{S{\rm max}} \rangle$, i.e., average velocity of the KDC; Column 5: average velocity in the region exterior to $\langle R_{S{\rm max}} \rangle$, i.e., average velocity of the main body of the dE;  Column 6: is the KDC a statistically significant detection based on the criterion described in the text?; Column 7: reference to the measurement of the rotation curve. Th06 for \citet{Thomas06}, Ch09 for \citet{Chil09}, and Ko11 for \citet{Koleva11}. See the text for details on the measurement of these parameters. \\
$^*$VCC~1036 contains a significant kinematic anomaly near its center. However, this galaxy is different from the other six galaxies discussed in this paper in that its kinematic anomaly does not resemble a symmetric KDC (see Appendix \ref{app}).

\end{table*}

\subsection{The Fraction of dEs that Contain KDCs}

There is a total of 101 available rotation curves for dEs in different environments combining Paper~II and the literature with similar or higher spectral resolution than ours \citep[][SAURON IFU spectra are considered here because, despite their lower spectral resolution, their two dimensional spatial coverage favors finding KDCs]{Ped02,Geha02,Geha03,VZ04,DR05,Thomas06,Chil09,etj11,Spolaor10,Koleva11,Rys13}. This number is large enough to statistically quantify how frequently KDCs are detected within this galaxy class.

The resulting statistics presented here must be used cautiously for four main reasons: (1) the velocity resolution plays against finding features with very low velocity amplitudes; (2) the majority of the rotation curves come from long-slit data and  thus are limited to those KDCs that are nearly aligned with the $PA$ of the slit; (3) the absence of significant evidence for the presence of a KDC does not rule out its existence; and (4) the sample of dEs observed spectroscopically is not complete.

Besides these technical problems for detecting KDCs, there is an observational bias that not even integral-field spectroscopy or larger surveys can overcome: the fading of the stellar populations over time. \citet{McDermid06} simulate this effect and find that in $2-5$~Gyr a KDC is no longer bright enough with respect to the underlying main body of the galaxy to be detectable.
These technical and observational biases indicate that the number of galaxies hosting a KDC could be larger than the number we currently know, although the rate at which KDCs are being formed is yet unknown.

The number of KDCs detected in dEs located in groups or lower density environments \citep[dEs without neighbors within a radius of $\sim 30'$, as defined by][]{Rys13} is $14.3\pm10.1\%$ \citep[2 out of 14 dEs surveyed by][]{DR05,etj11,Koleva11,Rys13}.
In the Fornax cluster, none of the 16 dEs analyzed exhibit a KDC \citep{DR05,Spolaor10,Koleva11}. 
In the Virgo cluster, $5.6\pm2.8\%$ ($8.5 \pm 3.4$, including those with marginal detections in Table \ref{tableKDCs}) have statistically reliable detections of a KDC. These are 4/71 (6/71 including marginal detections) of the dEs analyzed in this cluster \citep[][Paper~II]{Ped02,SimPrugVI,Geha02,Geha03,VZ04,Thomas06,Chil09,etj11,Spolaor10,Koleva11,Rys13}. This sample of 71 Virgo cluster dEs spans an $r$-band absolute magnitude range of $-19.3 < M_r < -16.1$ and represents $44\%$ of the early-type galaxies in the Virgo cluster with updated memberships by heliocentric velocities in that magnitude range \citep[71 out of the total 161,][]{Lisk07}.

Combining the numbers for statistically significant detections of KDCs in dEs located in low density environments, the Fornax, and the Virgo clusters, $5.9\pm2.4\%$ ($7.9 \pm 2.8$, including those with marginal detections) of the dEs host a KDC (or $5.1\pm0.8\%$ if we only count the 39 dEs homogeneously analyzed in this work). For luminous ETGs, $8.1\pm1.8\%$ host a KDC \citep[based on the ATLAS$^{\rm 3D}$ volume complete survey,][]{Em11,Kraj11}. Besides the caveats described above, the rate of occurrence of KDCs in dEs is very similar to the rate of occurrence of KDCs in ETGs. 

\section{Discussion of Possible Formation Scenarios} \label{disc}

The two most discussed mechanisms by which late-type galaxies are transformed into dEs --- ram-pressure stripping and harassment --- have typically been invoked to explain all the properties of dEs. The presence of KDCs in a small fraction of dEs adds a new twist to the story. 

\subsection{Insights from KDCs}

Ram-pressure stripping, hydrodynamical interaction of the dE progenitor with the hot intracluster medium, does not provide an obvious explanation for the formation of KDCs. This mechanism has no direct effect on the stars in the progenitor galaxy. Moreover, while the inflow of cold gas could produce a KDC, the ram-pressure stripping mechanism is more likely to lead to the expulsion rather than inflow of gas \citep[e.g.,][]{LinFab83,Boselli08a}. Harassment, multiple gravitational interactions of the dE progenitor with other cluster members, can produce large-scale counter-rotating structures ($R >1.5R_e$) but not a KDC \citep[][$R \ll R_e$]{GonzalezGarcia05}. An example of such a large-scale ($R \gtrsim R_e$) counter-rotating feature can be found in the close M31 dE satellite NGC~205 \citep{Geha06}. Despite its proximity to M31 (8~kpc in projection), the core of this dE shows no apparent signs of perturbation by M31's tidal forces.

In luminous ETGs, a KDC can be produced in a dissipationless merger (i.e., dry merger). The tightly bound core of one of the galaxies can become the center of the final object \citep[e.g.,][]{Kormendy84,Holley00}. However, the stellar populations of the central component are not necessarily younger or more metal rich than the main body of the galaxy, as it is found in the KDCs analyzed here.

KDCs that are younger than the main body of the galaxy and have diameter scales below $\sim 1$~kpc are always found in fast rotating luminous ETGs. These are likely formed from gas accretion that fall into the center of the galaxy \citep{McDermid06}. 
The properties of the KDCs found in VCC~1183 and VCC~1453 agree well with those found in fast rotating luminous ETGs. Moreover, VCC~1183 and VCC~1453 have ($\lambda_{Re/2},\epsilon_{1/2}$) parameters of ($0.14\pm0.04,0.102\pm0.002$) and ($0.15\pm0.03,0.188\pm0.002$), respectively, inside half of the $R_e$. These values place these two galaxies also as fast rotators\footnote{See Paper~III of this series for details on the measurement of $\lambda_{Re/2}$ and \citet{Em07,Em11} for the definition of $\lambda_R$ and slow/fast rotators.}. Thus, these features could be formed from gas accretion, but, how do two dEs in the core of the Virgo cluster \citep[located at $0.3R_{\rm virial}$ and $0.2R_{\rm virial}$, respectively, assuming the projected virial radius to be $R_{\rm virial}=5.38^{\circ}$,][]{Ferrarese12} accrete cold gas or hot gas that later cools down and forms stars?

Even though dwarf-dwarf mergers are not expected to be very common, they can occur \citep[e.g.,][]{deLucia06,Fakhouri10}. Contrary to what happens in massive galaxies, the structural parameters of the remnant dwarf are similar to those of its parents, i.e., the surface brightness profile remains similar to exponential \citep{Kazantzidis11}, in agreement with the structural parameters of dEs (see the S\'ersic index in Table \ref{props}). However, there are two main arguments against mergers happening inside a cluster: (1) the relative velocities of galaxies inside a cluster are too high to merge galaxies \citep[e.g.,][]{BinnTrem87,BG06}; and (2) the time to cool the gas down and form stars is significantly larger than the timescale to quench galaxies via ram-pressure stripping \citep[$\sim 150$ Myr,][]{Boselli08a}. Thus, dwarf-dwarf mergers, or just gas accretion, must have happened in a lower density environment, such as a group or a pair of galaxies, where the relative velocity of the dwarf galaxies is smaller and the environmental quenching is slower.

In this scenario, the KDC would be formed while the dwarf galaxy is still in the group. Afterwards the group would enter the Virgo cluster \citep[as shown by simulations, e.g.,][]{Gnedin03} and the significantly hotter environment will quench the star formation. However, there is no direct evidence that these six Virgo dEs, that host a KDC, are part of a group.

\subsection{The environment of each KDC host}

There is no easy way to identify companion(s) for these galaxies that were possibly part of a group before entering the cluster environment. In any case, not being able to find their companion(s) does not discard this scenario where KDCs are formed when the galaxies are part of groups or pairs that subsequently fell into the Virgo cluster. The survival of such groups and pairs as gravitationally bound structures of galaxies is not expected to be long inside a cluster environment.

We explore the surroundings for each Virgo cluster dE possibly hosting a KDC. We limit our study to neighbors within a projected distance of 100~kpc for whom the radial velocity is known.

\subsubsection{VCC~510}
VCC~510, with a radial velocity of 838 km~s$^{-1}$, is at a close projected distance to some galaxies in Virgo \citep[radial velocities from][unless otherwise indicated]{GOLDMine}. VCC~490 is a dE located at 33.1~kpc ($6.9'$) in projection from VCC~510 and has a velocity of 1267~km~s$^{-1}$. VCC~545, is a dE at 39.3~kpc ($8.3'$) with a radial velocity of 1207~km~s$^{-1}$. VCC~559 (NGC 4312) is a spiral galaxy (Sab) at 52.9~kpc ($11.12'$) with a radial velocity of 148~km~s$^{-1}$. Finally, VCC~596 (M100) is a Sc spiral at 86.2~kpc ($18.1'$) from VCC~510 with a radial velocity of 1575~km~s$^{-1}$. All these galaxies, but for VCC~559 and probably also VCC~596, have radial velocities that are close enough to that of VCC~510 to suggest that they are part of a group that fell into the Virgo cluster.

\subsubsection{VCC~917}
VCC~917 has a radial velocity of $1258.9 \pm 0.2$~km~s$^{-1}$ (Paper~II). This dE is located at 72~kpc ($15.1'$) from VCC~1001, an irregular galaxy with a radial velocity of 340~km~s$^{-1}$. The large difference between these two radial velocities suggests that they are not gravitationally bound. 

\subsubsection{VCC~1036}
The metric described in Section \ref{quantKDC} does not find a statistically significant KDC in this galaxy. However, its rotation curve is asymmetric which makes it interesting to explore its surroundings. 

VCC~1036 has a radial velocity of 1124~km~s$^{-1}$ and has three dEs close in projection whose radial velocities are similar to that of VCC~1036. This suggests that these dEs  could be part of the same gravitationally bound structure that fell into the Virgo cluster. The three dEs are: (1) VCC~940, $\sim 74$~kpc ($15.6'$) away from VCC~1036 with a radial velocity of $1410.5 \pm 0.4$~km~s$^{-1}$ (Paper~II); (2) VCC~965, $\sim 83$~kpc ($17.4'$) away from VCC~1036 with a radial velocity of 851~km~s$^{-1}$; and (3) VCC~1010, $\sim 17$~kpc ($3.7'$) away from VCC~1036 with a radial velocity of $915.5\pm0.3$~km~s$^{-1}$ (Paper~II).

In addition, VCC~1036 is also in close projection to VCC~1035 (a luminous elliptical) and VCC~1047 (a spiral galaxy). The radial velocities are very different from that of VCC~1036: $-500$~km~s$^{-1}$ and 724~km~s$^{-1}$, respectively. These two large galaxies may be responsible for some of the asymmetries observed in the rotation curve of VCC~1036.

\subsubsection{VCC~1183}
\begin{figure}
\centering
\includegraphics[angle=0,width=8.5cm]{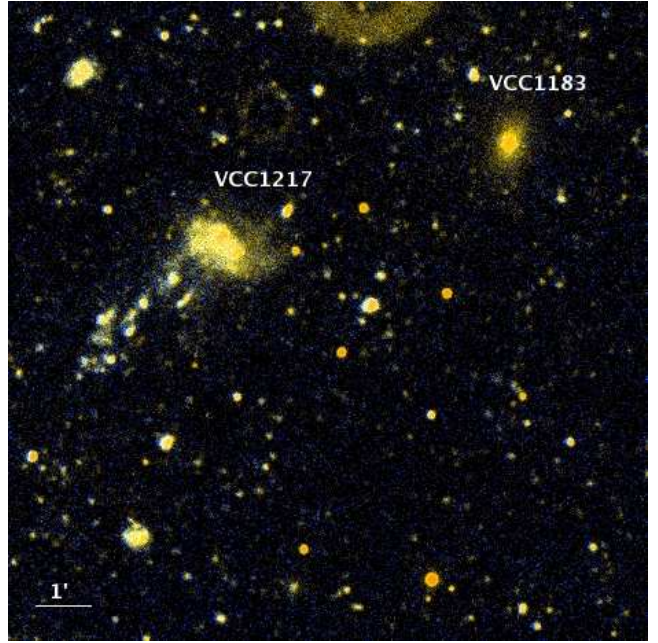}
\caption{GALEX UV color map of VCC~1183 (dE on the right) and its closest neighbor VCC~1217 (the dwarf star-forming galaxy with a tail of star-forming clumps on the left). Both galaxies are separated $6.6'$ ($\sim 22$~kpc). The idle semicircular orange artifact in the top edge of the image is caused by a nearby bright star.
\label{colormap}}
\end{figure}

VCC~1183 is in close projection to the most disturbed dwarf star-forming galaxy in Virgo, VCC~1217 \citep{Hester10}, as seen in Figure \ref{colormap}. This Figure shows a deep ultraviolet (UV) color map that combines the far UV and near UV channels of GALEX taken by the GUViCS collaboration \citep[GALEX Ultra Violet Virgo Cluster Survey,][]{Boselli11}.
Despite their apparent projected proximity ($\sim 22$~kpc apart), their radial velocities \citep[$1300.9 \pm 0.5$~km~s$^{-1}$ versus 176~km~s$^{-1}$, respectively --- this work;][]{Kenney13} suggest that they are unbound. Although we do not know the total three-dimension distance between these two galaxies or their true relative velocities, we carry out a simple calculation to constrain the age of the possible interaction between them. The minimum time $t$ since their last close passage is given by the quotient $r_{\rm min}/v_{\rm max}$, where $r_{\rm min}=22$~kpc and $v_{\rm max}$ is their maximum relative velocity. As these two galaxies are part of the Virgo cluster, their maximum relative velocity is $v_{\rm max}=3\sigma_{\rm Virgo}$, where the velocity dispersion of the Virgo cluster is $\sigma_{\rm Virgo}=753$~km~s$^{-1}$ \citep{BG06}. Thus, the last close passage occurred at least $t=10$ Myr ago. This timescale is neither consistent with the luminosity-weighted age of the stellar populations of the KDC (2~Gyr) nor with the age of the stellar populations of the tail of star-forming clumps (fireballs) observed in VCC~1217, which has been dated to be younger than 400 Myr \citep{Fumagalli11,Kenney13}. The energy available for an encounter between these two galaxies is very little because it scales as $\Delta V^{-2}$, where $\Delta V$ is their radial velocity difference \citep{BinnTrem87}. This energy is not enough to generate a KDC in VCC~1183 or a tail of fireballs in VCC~1217. In addition, there are several evidence, such as the structural features and lack of old stars in VCC~1217's tail, that point ram-pressure stripping as the mechanism responsible for the fireballs on VCC~1217 \citep{Kenney13}. 

\subsubsection{VCC~1261}
VCC~1261 has a radial velocity of $1799.4 \pm 0.3$~km~s$^{-1}$ (Paper~II) and has two close neighbors in projection. VCC~1200  is an irregular galaxy located at $\sim 42$~kpc ($9'$) from VCC~1261 and has a radial velocity of $-123$~km~s$^{-1}$. VCC~1269 is a dE located at $\sim 71$~kpc ($15'$) and has a radial velocity of 633~km~s$^{-1}$. The large velocity differences among these three galaxies indicate that they are not gravitationally bound to one another in the present time.

\subsubsection{VCC~1453}
VCC~1453 is located in projected distance at $\sim 58$~kpc ($17'$) from VCC~1401 (M88), a large spiral galaxy. The radial velocities of these two galaxies are $1859.2 \pm 0.3$~km~s$^{-1}$ (this work), and 2282~km~s$^{-1}$, respectively. These conditions suggest that these two galaxies are gravitationally bound to each other implying that they could have been part of the same group that fell into the Virgo cluster. 

\subsubsection{VCC~1475}
VCC~1475 has a radial velocity of 951~km~s$^{-1}$. This bright dE \citep[$M_r=-19.0$,][]{Janz09} is close in projection (85.3~kpc, $17.9'$) to VCC~1427, an Im/blue compact dwarf with a radial velocity of $-132$~km~s$^{-1}$. The large difference between the radial velocities of these galaxies suggests that they are not gravitationally bound.

\section{Summary and conclusions} \label{concl}

In this work, we present the discovery of kinematically decoupled cores in the Virgo cluster dEs VCC~1183 and VCC~1453. Combined with the KDCs found in VCC~510 by \citet{Thomas06}, VCC~917 and VCC~1261 by \citet{Chil09}, and in VCC~1475 by \citet{Koleva11}, this makes a total of six dEs (defined as early-type galaxies in the luminosity range $-19 < M_r < -16$) with a significant or marginal detection of a KDC in the Virgo cluster. For the two new KDCs presented in this paper, we provide radial profiles of the rotation, velocity dispersion, and stellar population indices. These spectroscopic measurements are also compared with deep photometry in the $H$-band.

The KDCs for VCC~1183 and VCC~1453 have radii of 1.8$''$ (0.14~kpc), and 4.2$''$ (0.33~kpc), respectively. The stellar populations of the KDCs are younger and possibly more metal rich than the main body of the galaxy. For VCC~1183, the KDC is $2.0\pm0.5$~Gyr old and has a metallicity of ${\rm [M/H]} = -0.3\pm0.3$, while the main body of the galaxy is $9.9\pm4.1$~Gyr old and has a metallicity of  ${\rm [M/H]} = -0.6\pm0.3$. For VCC~1453, the KDC is $4.5\pm1.1$~Gyr old with a metallicity of ${\rm [M/H]} = -0.2\pm0.1$, and the main body is $\gtrsim 10$~Gyr old with a metallicity of ${\rm [M/H]} = -1.1\pm0.3$. These stellar populations for the main bodies are similar to the typical values of other dEs in the Virgo cluster \citep{Mich08,etj09,Paudel11}. This difference in the measured stellar populations between the KDC and further out the main body could also be related to an intrinsic gradient in the host galaxy.

Photometrically, there does not appear to be a strong correlation between any particular feature and the KDC. VCC~1183 seems to be very complex structurally. At $R \sim 6''$ the isophotes are twisted $\sim 77^{\circ}$ and the ellipticity drops with respect to the outer parts of the galaxy. This region covers our full kinematical measurements including the KDC and the main body of the galaxy.

The formation scenarios that seem more plausible to explain the properties found here, involve dwarf-dwarf wet mergers and gas accretion. Due to the density of the Virgo cluster, such events can only happen in lower density environments such as poor groups or galaxy pairs.

\acknowledgments

E.T. thanks Laura Ferrarese for very useful suggestions on ellipse-fitting and Eric Emsellem, Jeffrey Kenney, and Marla Geha for very helpful discussions. The authors thank the anonymous referee for useful suggestions that helped improve the manuscript.
E.T. acknowledges the financial support of the Fulbright Program jointly with the Spanish Ministry of Education. PG acknowledges the NSF grant AST-1010039. GvdV and JFB acknowledge the DAGAL network from the People Programme (Marie Curie Actions) of the European Union’s Seventh Framework Programme FP7/2007-2013/ under REA grant agreement number PITN-GA-2011-289313. T.L. was supported within the framework of the Excellence Initiative by the German Research Foundation (DFG) through the Heidelberg Graduate School of Fundamental Physics (grant number GSC 129/1). This work has made use of the GOLDMine database \citep{GOLDMine} and the NASA/IPAC Extragalactic Database (NED.\footnote{http://ned.ipac.caltech.edu})

\bibliographystyle{aa}
\bibliography{references}{}

\appendix

\section{Folded Rotation Curves and Statistical Significance of KDCs}\label{app}

\begin{figure}
\centering
\includegraphics[angle=-90,width=8cm]{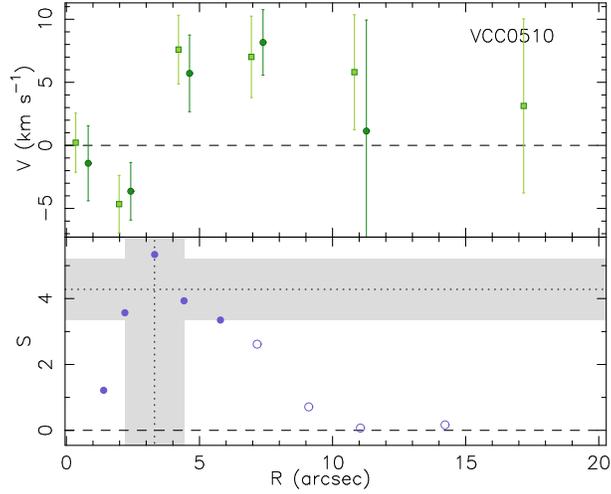}
\caption{{\bf Upper panel:} Folded rotation curve for VCC~510. The light green squares and dark green dots indicate opposite sides of the rotation curve with respect to the center of the galaxy. {\bf Lower pannel:} Statistical parameter $S(R)$. See Section \ref{quantKDC1} for its definition. Filled purple circles indicate that $\langle V_{\rm inner}(<R) \rangle$ and $\langle V_{\rm outer}(>R) \rangle$ have opposite signs. Open purple circles indicate that $\langle V_{\rm inner}(<R) \rangle$ and $\langle V_{\rm outer}(>R) \rangle$ have the same sign. The vertical and horizontal dotted lines show the values for $\langle R_{S{\rm max}} \rangle$ and $\langle S_{{\rm max}} \rangle$. The shaded grey areas indicate the uncertainties in  $\langle R_{S{\rm max}} \rangle$ and $\langle S_{{\rm max}} \rangle$, respectively.
\label{appVCC0510}}
\end{figure}

\begin{figure}
\centering
\includegraphics[angle=-90,width=8cm]{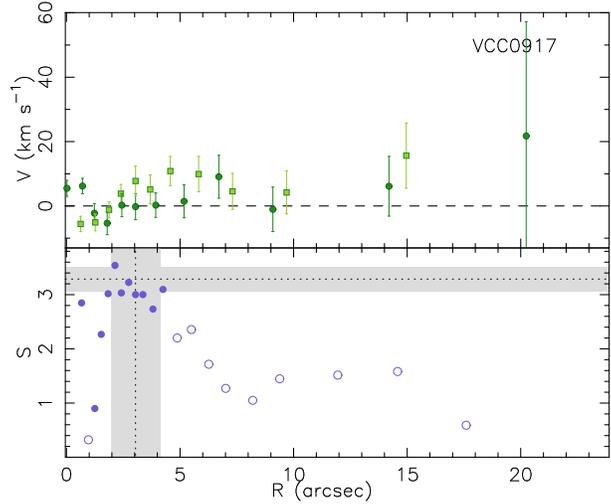}
\caption{Same as Figure \ref{appVCC0510} for VCC~917.
\label{appVCC0917}}
\end{figure}

\begin{figure}
\centering
\includegraphics[angle=-90,width=8cm]{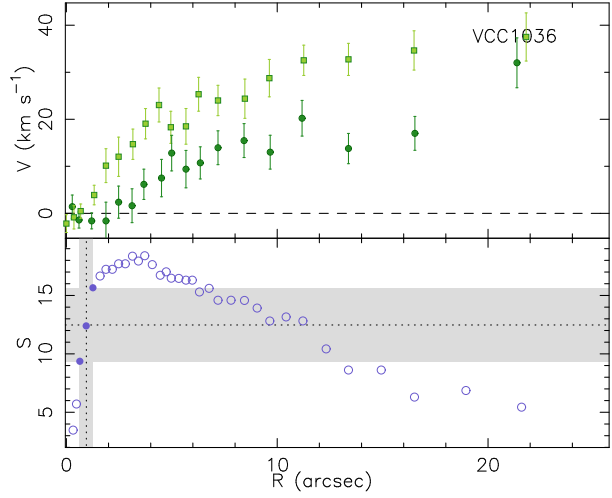}
\caption{Same as Figure \ref{appVCC0510} for VCC~1036. This galaxy shows a significant kinematic anomaly near its center that does not resemble a symmetric KDC (see the text for details).
\label{appVCC1036}}
\end{figure}

\begin{figure}
\centering
\includegraphics[angle=-90,width=8cm]{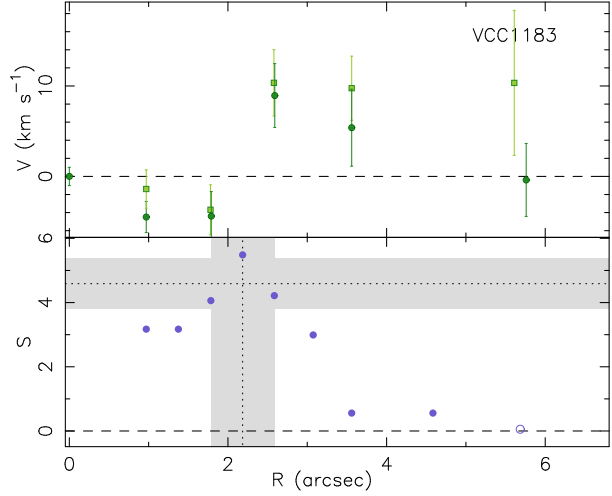}
\caption{Same as Figure \ref{appVCC0510} for VCC~1183.
\label{appVCC1183}}
\end{figure}

\begin{figure}
\centering
\includegraphics[angle=-90,width=8cm]{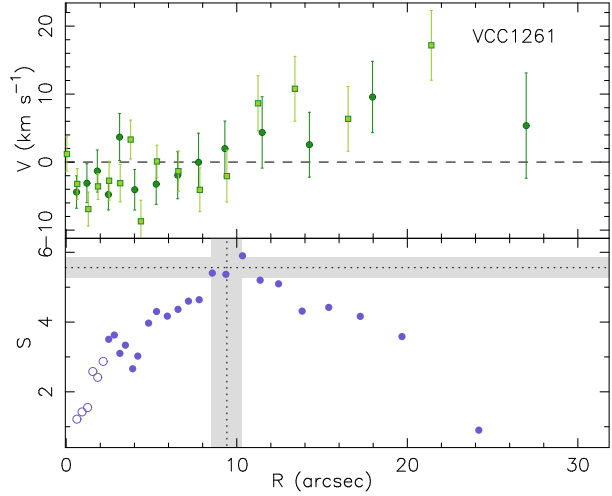}
\caption{Same as Figure \ref{appVCC0510} for VCC~1261.
\label{appVCC1261}}
\end{figure}

\begin{figure}
\centering
\includegraphics[angle=-90,width=8cm]{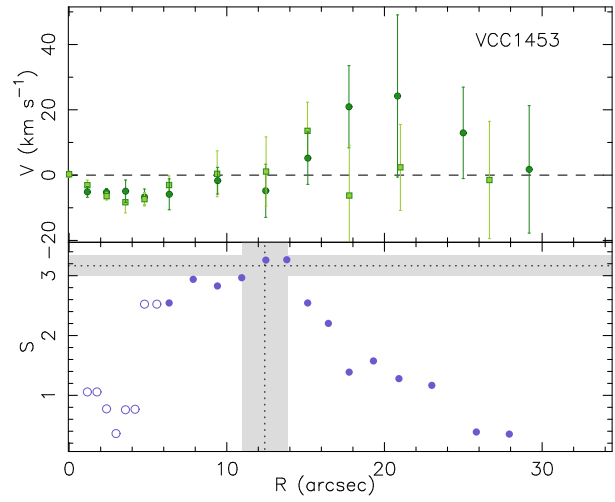}
\caption{Same as Figure \ref{appVCC0510} for VCC~1453.
\label{appVCC1453}}
\end{figure}

\begin{figure}
\centering
\includegraphics[angle=-90,width=8cm]{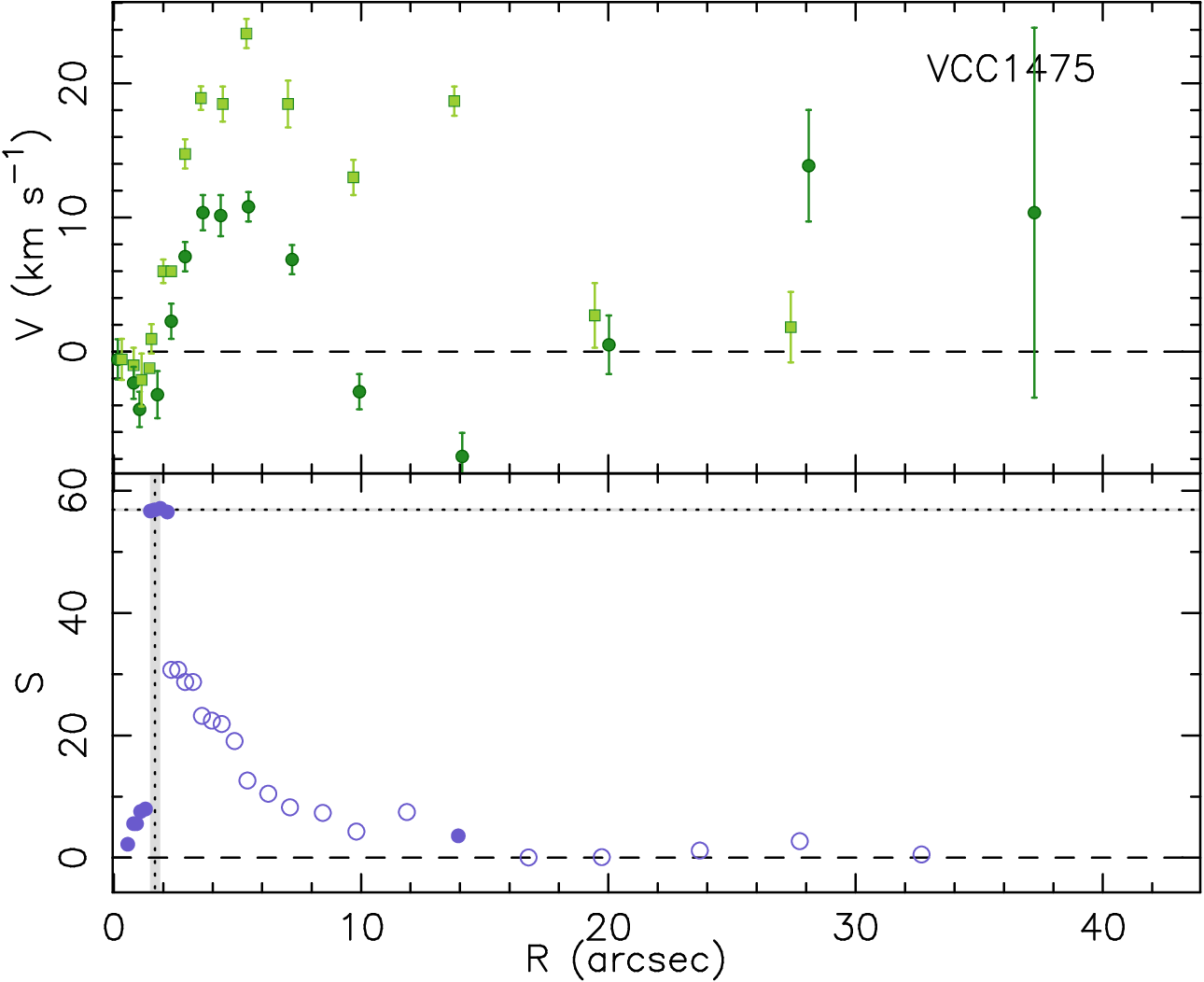}
\caption{Same as Figure \ref{appVCC0510} for VCC~1475.
\label{appVCC1475}}
\end{figure}

Figures \ref{appVCC0510} to \ref{appVCC1475} illustrate the statistical metric described in Section \ref{quantKDC}. Each figure shows the folded rotation curve for a dE possibly hosting a KDC. The statistical parameter $S(R)$ defined in Section \ref{quantKDC1} is also shown for each galaxy. The maximum of the statistical parameter $S(R)$ indicates the significance of the detection of the KDC. When $\langle S_{\rm max} \rangle \gg 3$ the detection of the KDC is significant, when $\langle S_{\rm max} \rangle \sim 3$ the detection is marginal. 

To distinguish between a monotonically increasing smooth rotation curve from a rotation curve with a KDC in its center, the maximum value of $S(R)$ is calculated in the region where $\langle V_{\rm inner}(<R) \rangle$ and $\langle V_{\rm outer}(>R) \rangle$ have opposite signs (filled purple circles in the lower pannels of Figures \ref{appVCC0510} to \ref{appVCC1475}). The parameter $\langle S_{\rm max} \rangle$ is the average of the three maximum values of $S(R)$ in that region and $\langle R_{S{\rm max}} \rangle$ is the average radius of the three maximum values of $S(R)$.

The $S(R)$ profile for VCC~1036 is different from the rest of the galaxies. The region where $\langle V_{\rm inner}(<R) \rangle$ and $\langle V_{\rm outer}(>R) \rangle$ have opposite signs do not show a peak in $S(R)$. The peak is in the region where $\langle V_{\rm inner}(<R) \rangle$ and $\langle V_{\rm outer}(>R) \rangle$ have the same sign. However,  $\langle S_{\rm max} \rangle \ggg 3$ which indicates that there is a significant kinematic anomaly near the center of this galaxy. But, that anomaly does not resemble a symmetric KDC.

\end{document}